
\input harvmac
\newcount\figno
\figno=0
\def\fig#1#2#3{
\par\begingroup\parindent=0pt\leftskip=1cm\rightskip=1cm\parindent=0pt
\baselineskip=11pt
\global\advance\figno by 1
\midinsert
\epsfxsize=#3
\centerline{\epsfbox{#2}}
\vskip 12pt
{\bf Fig. \the\figno:} #1\par
\endinsert\endgroup\par
}
\def\figlabel#1{\xdef#1{\the\figno}}
\def\encadremath#1{\vbox{\hrule\hbox{\vrule\kern8pt\vbox{\kern8pt
\hbox{$\displaystyle #1$}\kern8pt}
\kern8pt\vrule}\hrule}}

\overfullrule=0pt

%
\def\tilde{\widetilde}
\def\bar{\overline}

\font\zfont = cmss10 

\def\bigone{\hbox{1\kern -.23em {\rm l}}}
\def\ZZ{\hbox{\zfont Z\kern-.4emZ}}

\Title{hep-th/9505053, HUTP-95-A015, IASSNS-HEP-95-33}
{\vbox{\centerline{A One-Loop Test of String Duality}}}
\smallskip
\centerline{Cumrun Vafa}
\smallskip
\centerline{\it Lyman Laboratory of Physics, Harvard University}
\centerline{\it Cambridge, MA 02138 USA}
\smallskip
\centerline{and}
\smallskip
\centerline{Edward Witten}
\smallskip
\centerline{\it School of Natural Sciences, Institute for Advanced Study}
\centerline{\it Olden Lane, Princeton, NJ 08540, USA}\bigskip
\baselineskip 18pt

\medskip
\vskip 1cm
\noindent
We test type IIA-heterotic string duality in six dimensions
by showing that the sigma model anomaly of the  heterotic
string is generated by a combination of a tree level and a string one-loop
correction on the type IIA side.

\Date{May, 1995}

\newsec{Introduction}

Six dimensional string
theories with non-chiral $N=4$ supersymmetry can be constructed
most easily either by compactifying the heterotic string on a four-torus
or by compactifying a Type IIA superstring on a K3 manifold.
According to results of Narain \ref\narain{K. Narain,``New Heterotic
String Theories in Uncomapctified Dimensions $<10$,'' Phys. Lett.
{\bf 169 B} (1986) 41; K. Narain, M. Sarmadi, and E. Witten,``
A Note On the Toroidal Compactification of Heterotic String Theory,''
Phys. Rev. {\bf D35} (1987) 369.  }\
for the heterotic case and of Seiberg \ref\seib{N. Seiberg
``Observations On the Moduli Space of Superconformal Field
Theories,'' Nucl. Phys. {\bf B303}(1988) 286.}
and Aspinwall and Morrison \ref\aspmor{P. Aspinwall and D. Morrison,
``String Theory on K3 Surfaces,'' DUK-TH-94-68,IASSNS-HEP-94/23.}
for the Type IIA case, these theories have the same moduli spaces
of vacua.
(Globally, this result depends on mirror symmetry, as was shown
in \aspmor, giving an elegant illustration of how conformal
field theory behaves in a way that makes space-time stringy dualities
possible.)
\nref\hullt{C.M. Hull and P.K.  Townsend, ``Unity of Superstring
Dualities'', QMW-94-30, R/94/33.}
\nref\vafc{C. Vafa, unpublished.}
This hints that the two theories might be equivalent, as has been conjectured
\refs{\hullt,\vafc}.

\nref\ewit{E. Witten, ``String Theory Dynamics In Various Dimensions,''
hepth-9503124, to appear in Nucl. Phys. B.}
\nref\sen{A. Sen, ``String-String Duality
Conjecture in Six Dimensions and
Charged Solitonic Strings'', hep-th/9504027.}
\nref\harstrom{J.A. Harvey and A. Strominger, ``The
Heterotic String is a Soliton'', hep-th/9504047.}
\nref\strom{A. Strominger,
``Massless Black Holes and Conifolds in String Theory'', hep-th/950490;
B. Greene, D. Morrison and A. Strominger, ``Black Hole Condensation
and the Unification of String Vacua'', hep-th/9504145.}
\nref\vafa{C. Vafa,``A Stringy Test of the Fate of the Conifold'',
hep-th/9505023.}
The hint might appear unconvincing since the equivalence in question
is largely determined by the low energy supergravity.
However, the equivalence between
these theories, which has been called
 string-string duality, has been supported by a variety of new arguments
\refs{\ewit - \harstrom}.
Also, one of the strangest requirements of the six-dimensional
string-string duality, which is that the Type IIA theory must develop
massless charged black holes precisely when a cycle in the K3 collapses
(see section 4.6 in \ewit) has become much more plausible because
of dramatic results in four dimensions \strom\ that depend on an analogous
phenomenon.  These latter results have been further supported by
certain one-loop string calculations \vafa .

The purpose of the present paper is to subject six-dimensional
string-string duality to one small further test.  Compactification
of the heterotic string on a four-torus gives a six-dimensional
theory in which the allowed topology (of the manifold and gauge bundle)
is subject to certain restrictions.  We would like to verify that the
same restrictions also hold in the corresponding Type IIA theory.

One restriction is that the six-manifold must be a spin manifold,
since the heterotic string
has fermions. This restriction obviously holds, for the same reason,
also for the Type IIA theory.

The other known restriction comes from an equation that plays an
important role in anomaly cancellation.  If $H$ is the field strength
of the two-form, $F$ the Yang-Mills field strength, $R$ the Riemannian
curvature two-form, and $\tr$ the trace in the fundamental representation
(of an $SO(32)$ gauge group or of the Lorentz group $SO(9,1)$)
\foot{Our gauge bosons are real, antisymmetric matrices, as is natural
for $SO(N)$, so the trace is {\it negative} definite.} then
in the 10 dimensional heterotic string we have
\eqn\firstone{d H = -\tr F\wedge F+\tr R\wedge R.}
Upon toroidal compactification of heterotic strings to lower
dimensions we get extra gauge symmetries from the left-movers
and right-movers; the rank goes up two with each
compactified direction.  The extra gauge bosons appear in the six-dimensional
version of \firstone.

{}From \firstone\ it follows that the cohomology class represented by
$\tr F\wedge F -\tr R\wedge R$ is zero.  This cohomology class is,
roughly speaking, $p_1(V)-p_1(T)$, where $V$ and $T$ are the gauge and
tangent bundles and $p_1$ denotes the first Pontryagin class.
Actually, a more precise analysis
\ref\oldwit{E. Witten, ``Global Anomalies In String Theory,'' in
{\it Geometry, Anomalies, And Topology}, ed W. A. Bardeen and A. R. White
(World-Scientific, 1985).} including world-sheet global anomalies
shows that this condition can be imposed at the level of {\it integral}
cohomology, not just de Rham cohomology.  Also, for a real orientable vector
bundle whose structure group can be lifted to the spin group
(this is so for $T$ because fermions exist, and for $V$ because the
heterotic string contains gauge spinors) the first Pontryagin class
is divisible by 2 in a natural way.  The integral characteristic class obtained
by dividing it by 2 seems to have no standard name; we will simply call
it ${1\over 2}p_1$.  At any rate, the topological condition in the heterotic
string is really
\eqn\ippo{{p_1(V)\over 2}-{p_1(T)\over 2}=0}
with the classes understood as integral classes.

In the present paper, we will seek a condition similar to \ippo\ for the
Type IIA superstring theory compactified on K3.  Our analysis will not be
precise enough to see the torsion (we comment on this below),
but we will see the 2 in the denominator.
It is fairly obvious how we must proceed.
Since the relation between the field strength $H$ in the heterotic
string and the corresponding field strength $H'$ for Type IIA is
$H= *e^{-2\phi '}H'$ with $\phi'$ being the dilaton (the potential
importance of such a relation was foreseen by Duff and Minasian
\ref\duff{M.J. Duff
 and R. Minasian,``Putting String/String Duality to the Test,'' Nucl.
Phys. {\bf B436} (1995) 507.}),
we must replace \firstone\ by a relation of the form
\eqn\yippo{d^*(e^{-2\phi'}H')=-\tr F\wedge F+\tr R\wedge R.}
This will have to be the equation of motion of the two-form field
$B'$, so the six-dimensional effective Lagrangian must contain a term
with the structure
\eqn\hippo{\int B'\wedge \left(\tr F\wedge F-\tr R\wedge R\right).}
Concretely, if this interaction is present then the desired restriction
on the topology of the manifold and gauge bundle will hold because the
equations of motion \yippo\ have no solution otherwise.

The rest of this paper is devoted to finding the interactions just
claimed.  It is fairly obvious that the $\tr F\wedge F$ term must
be present at tree level as it is required by low energy supersymmetry.
(This term is present since it is dual to the $\tr F\wedge F$ term
in \firstone, which is likewise
required by low energy supergravity and so present
at tree level.)  We explain the details in section 2.  The $\tr R\wedge R$
term is perhaps more mysterious; it arises from a one-loop computation,
as we explain in section 3, which
gives a term $\int B'\wedge Y^8$ where
$Y^8$ is a characteristic class involving the Riemann tensor.
Compactifying upon K3 leads to a term of the form $\int B'\wedge \tr
R\wedge R$ as is required by the string duality.
Note that the presence of this correction at one loop implies
an inconsistency in type IIA string
compactifications to two dimensions on an eight-manifold with $\int Y^8\not=
0$; at the one loop level there is no extremum of the effective
action.
\foot{Type IIB compactification on the same
manifolds would be inconsistent if we compactify further on a circle
because then there would be no distinction between
Type IIA and IIB.
Also, a similar one loop term which exists for the heterotic string
and is responsible for the Green-Schwarz anomaly cancellation mechanism
would destabilize compactifications down to two  dimensions for which
this class is not zero.} For example if the eight-manifold is
${\rm K3}\times {\rm K3}$ the type IIA compactification is destabilized at
string one-loop.  The corresponding statement for the heterotic
side is that if we compactify heterotic string on $T^4\times {\rm K3}$
without turning on any gauge fields, the heterotic string is inconsistent
due to sigma model anomalies.  The way to remedy this problem on both
sides is to turn on an appropriate gauge field.

One might wonder how one could understand the torsion part of
\ippo\ for Type II superstrings.  Since duality tends to exchange
world-sheet and space-time effects, and the torsion shows up in world-sheet
global anomalies in the case of the heterotic string, one might look
for space-time global anomalies in Type II that would explain
the necessity for the torsion part of \ippo\ to vanish.  Sometimes
global anomalies are more obvious in a soliton or instanton sector
than in the vacuum sector.  One might in fact look to the solitonic
heterotic string of the Type II theory
\refs{\sen,\harstrom} as an object whose
quantization (like that of the elementary heterotic string) may require
vanishing of the torsion part of \ippo.

We must confess to a sin of omission: we have  not been precise with
orientation conventions and so have not checked the relative sign of
the $\tr F\wedge F $ and $\tr R\wedge R$ interactions.

We understand that some of the issues in this paper have also been
considered in unpublished work by J. Harvey.

\newsec{The $\tr F\wedge F$ Interaction   }

In what follows, we will
deduce from ten-dimensional Type IIA supergravity
the $B'\wedge \tr F\wedge F$ interaction.  Since the existence of
this interaction is certainly already known,
the only slightly non-trivial point is to determine the correct
normalization; for this we will need a recent result by Harvey and
Strominger \harstrom.
Also, in the rest of this paper, we consider
only the Type IIA superstring theory, and relabel $B'$ simply as $B$.

In the conventions of
\ref\othercal{C. G.
Callan, Jr., J. A. Harvey, and A. Strominger, ``Worldsheet Approach
To Heterotic Instantons And Solitons,'' Nucl. Phys. {\bf B359} (1991) 611.}
(which are used in \harstrom), the world-sheet coupling involving
the $B$ field is
\eqn\hobo{L_B={1\over 4\pi\alpha'}\int_\Sigma\epsilon^{ab}B_{IJ}\partial_a
X^I\partial_b X^J.}
It follows, in particular, that the period of the $B$ field is $4\pi^2\alpha'$.
This means that $L_B$ is invariant, modulo $2\pi$ times an integer,
under the addition to $B$ of a closed form $\beta$ with the property
that for every closed surface $C$ in the target space,
\eqn\nobo{\int_C \beta}
is an integer multiple of $4\pi^2\alpha'$.
\foot{To be very precise about the meaning of \nobo, if $C$ is the
quotient of the $x^1-x^2$ plane by $x^1\to x^1+1,\,x^2\to x^2+1$,
and $\beta $ has non-zero values $\beta_{12}=-\beta_{21}=1$,
then $\int_C\beta=1$.}  We will call a shift $B\to B+\beta$ with such $\beta$
a global world-sheet gauge transformation.

Ten-dimensional Type IIA supergravity contains a three-form $C$ with field
strength $G$ -- in components $G_{IJKL}=\partial_IC_{JKL}\pm \dots$.
(We will reserve the name $F$ for the gauge field strengths that will
soon appear in six dimensions.)
This field couples to $B$ with a coupling of the general form
$B\wedge G\wedge G$.
Harvey and Strominger normalize $C$ so that this coupling (in a space-time
of Lorentz signature) is
\eqn\nurky{-{1\over 4\pi\alpha'{}^3}\int B\wedge G\wedge G.}
With this normalization, they argue that periods of $G$ are quantized
to be integral multiples of $\alpha'$.  In other words, if
$\Sigma^4$ is any closed four-surface in space-time, then
\eqn\hurky{\int_{\Sigma^4} G=n\alpha',~~~{\rm with}~n\in {\bf Z}.}
(For the normalization of such an integral, see the footnote above.)
Note that as $B$ has periods $4\pi^2\alpha'$, the existence of the
interaction \nurky\ implies that $G\wedge G$ must be $2\alpha'^2$
times an integral class; \hurky\ implies the slightly weaker result
that $G\wedge G$ is $\alpha'^2$ times an integral class. In compactification
to six dimensions on a spin manifold, we will gain an extra factor
of two from the fact that the intersection form on the two-dimensional
cohomology will be even.
The meaning of the factor of 2 in the uncompactified ten-dimensional
theory is not clear.

Now, consider the compactification of the ten-dimensional theory
to six dimensions on a K3 manifold $X$.  Let $U_I$, $I=1\dots 24$,
be a basis of harmonic two-forms on $X$ with integral periods.  Then
\eqn\gurky{\int_XU_I\wedge U_J=d_{IJ},}
where $d_{IJ}$, the intersection pairing of K3, is even and unimodular
with signature $(4,20)$ (four positive and twenty negative eigenvalues).
Dimensional reduction
of the $C$ field to six dimensions is made by writing
\eqn\burky{C={\alpha'\over 2\pi}\sum_IA^I\wedge U_I,}
where the $A^I$ are $U(1)$ gauge fields in six dimensions.
The factor of $\alpha'\over 2\pi$ in \burky\ is chosen so that, in
view of \hurky, the field strengths $F^I=dA^I$ obey a conventionally
normalized Dirac condition
\eqn\lurky{\int_{\Sigma_2}F^I\in 2\pi{\bf Z}.}
This condition means that the $A^I$ can be interpreted as $U(1)$ gauge
fields, coupled to integral charges.

Via \burky, \nurky\ reduces in six dimensions to
\eqn\turky{L_1=
-{1\over 4\pi\alpha'}
\int B\wedge \sum_{I,J}{d_{IJ}F^I\wedge F^J\over (2\pi)^2}.}
Now, let us verify that (i) $L_1$ has the correct periodicity with
respect to a global world-sheet gauge transformation; (ii)
$L_1$ is minimal in the sense that $L_1/n$, with $n$ an integer $>1$,
would not have the right periodicity.
Under $B\to B+\beta$, $L_1$ changes by
\eqn\urky{\Delta L_1=-\pi \int {\beta\over 4\pi^2\alpha'}\wedge
\sum_{IJ}{d_{IJ}F^I\wedge F^J\over (2\pi)^2}.}
This has the properties claimed because $\beta/4\pi^2\alpha'$ is
an arbitrary closed form with integral periods, the $F^I/2\pi$ are arbitrary
forms with integral
periods, and a factor of 2 comes from the fact that $d$ is even;
thus $\Delta L_1$ is an integral multiple of $2\pi$ but not necessarily an
integral model of $2\pi n$ for any $n>1$.

To express this in terms of the first Pontryagin class, we can split
off a sixteen-dimensional sublattice of $H^2(X,{\bf Z})$ on which
the intersection form is equivalent to $-1$ times the usual
unimodular, integral form on the weight lattice of $Spin(32)/{\bf Z}_2$.
(One could similarly use $E_8\times E_8$, of course.)  If then
we interpret the $F^I$ (restricted to the sixteen-dimensional subspace)
as the ``Cartan'' part of an $SO(32)$ gauge field, then
$d_{IJ}F^I\wedge F^J$ can be identified with ${1\over 2}\tr F\wedge F$,
the trace now being the trace in the vector representation of $SO(32)$.
We can then rewrite
\turky\ in the form
\eqn\uturky{L_1=
-{1\over 2\pi\alpha'}
\int B\wedge {\tr F\wedge F\over 16\pi^2}.}
Here
\eqn\iturky{\Theta_V={\tr F\wedge F\over 16\pi^2}}
represents the first Pontryagin class $p_1(V)$.

If we write
\eqn\ituryky{\tilde B={B\over 4\pi^2 \alpha'}}
-- so that $\tilde B$ has integer periods -- then \iturky\ becomes
\eqn\yturky{L_1=-2\pi\int \tilde B\wedge{\Theta_V\over 2}.}
This has the right periodicity (it shifts by an integer multiple of $2\pi$
under global gauge transformations of $\tilde B$) because in
$Spin(32)/{\bf Z}_2$, the differential form $\Theta_V/2$ has integral
periods; it represents the class $p_1(V)/2$ which as mentioned in the
introduction is an integral class for bundles that admit spinors.

The discussion so far has been carried out in Lorentz signature.
Since the integrand in the Feynman path integral in Lorentz signature
is $e^{iL}$, while in Euclidean signature it is $e^{-L_E}$, the relation
between the two (for an interaction such as \yturky\ that is independent
of the metric and so does not explicitly ``see'' the signature) is
$L_E=-iL$, so in Euclidean signature our interaction would be
\eqn\gturky{L_1^E=2\pi i\int \tilde B\wedge {\Theta_V\over 2}.}

\newsec{The $\tr R\wedge R$ Interaction   }

This section is devoted to finding the six-dimensional $B\wedge \tr R\wedge R$
interaction whose necessity was explained in the introduction.
We first show that in ten dimensional Type IIA, there is a one-loop
contribution to the effective action of the form
\eqn\cor{\delta S=\int B Y_8}
where $Y_8$ is an eight dimensional characteristic
class made of the Riemann tensor contracted with one $\epsilon$
tensor. The desired six-dimensional interaction then follows upon
compactification on K3.

The computation that gives \cor\ is quite similar to
familiar computations of anomaly cancellation for the heterotic string
\ref\lsw{W. Lerche, B.E.W. Nilsson, A.N. Schellekens and N.P. Warner, ``
Anomaly Cancelling Terms From The Elliptic Genus'',
 Nucl. Phys. {\bf B299} (1988) 91. }.
The novelty here is
that a somewhat similar term arises for the Type IIA
superstring even though this theory is
non-chiral.  (There is no such term for the chiral Type IIB
theory.)
The computational methods of \lsw\ actually
directly apply. However for the sake
of completeness we will do the computation in our specific case, taking
a shortcut in obtaining the final answer.

 There are two
ways to do the computation:  one may, as in \lsw, compute directly
in 10 dimensions the one-loop
amplitude involving 4 gravitons and one anti-symmetric
tensor field and extract the piece which has
the correct index structure; or one may compactify on an eight
dimensional manifold $M$ all the way down to two dimensions and compute
the 1 point function of the $B$ field.  The constant of proportionality
will be a characteristic class of $M$ which can be rewritten in terms
of the Riemann tensor.  The first method is more direct
but more difficult.  We thus use
the second method; applied to the heterotic case this would yield
the results already computed in \lsw .

For Type II strings, one must choose independently
even or odd spin structures for left- and right-movers.
The $\epsilon$ tensor in \cor\ will arise
in worldsheet computations
from the absorption of fermion zero modes, which
appear when we have the odd spin-structure.  So
a term of the form \cor\ can only be generated
from computations where the left- or right-movers, but not both,
are in an odd spin structure.

In the Type IIB computation
the two contributions coming from (even,odd) and (odd,even) cancel
out by symmetry, whereas they add in the Type IIA computation.
The statements follow from the following.
The Type IIA theory is invariant under a parity transformation in space-time
combined with a parity transformation of the world-sheet.
(To make the Type IIA theory, one uses opposite GSO projections for
the left- and right-movers, leading to space-time spinors of opposite
chirality.  A space-time parity transformation, which exchanges the two
types of spinor, must thus be combined with a world-sheet parity
transformation.)  The interaction \cor\
contains an $\epsilon$ tensor, which  is odd under space-time parity,
and a  $B$ field, which is odd under world-sheet parity;
altogether \cor\ respects the symmetry of the Type IIA theory.
By contrast, the Type IIB theory  is invariant under world-sheet
parity (unaccompanied by any space-time transformation); this forbids
the interaction \cor, which is odd under $B\to -B$.

\bigskip\noindent{\it The Computation}

For the left- or right-moving sector with an odd spin structure
there is a supermodulus, which means that we have to use
the $-1$ picture for one of the vertex operators, and in addition
insert the supercurrent $G$ which comes from integration over
supermoduli.  On the side with even spin structure
we will use the picture 0 for the vertex operator.  Let
us take the left-moving fermions to be in the odd spin structure and the
right-moving fermions to be in the even spin structure.
 The vertex operator for the $B$-field which is
in picture $(-1,0)$ for (left, right) side is
$$V_B=i\delta(\gamma )B_{\mu \nu}\psi^\mu \bar \partial X^{\nu}$$
where we are only interested in the zero momentum contribution
so we have set $k=0$. This is the vertex operator normalized
so that $B\in H^2({\bf Z})$, i.e. has integral periodicity.  (This
object was called $\tilde B$ in the last section.)
  Note also that the left moving
supersymmetry generator on the worldsheet is
$$G=\psi^\alpha \partial X^\alpha$$
The one loop partition function is obtained
by integration over the fundamental domain of
$${-1\over 4}\int_{\cal M}\langle b(\mu)\bar b (\bar \mu)
[G \delta(\beta )] \ \bar c cV_B \rangle$$
The $1/4$ factor in front comes from various contributions:
a ${1/2}$ from $Z_2$ symmetry of torus, a factor of ${1\over 4}$
from the GSO projection and a factor of 2 because the
(odd, even) spin structures would give the same result as
(even, odd).  Here
 $b(\mu)$ is $b$ folded with the Beltrami differential
for the torus, and ${\cal M}$ is the moduli space of a torus with an
(even,odd) spin structure pair (this is simply three copies of
the fundamental domain without spin structures).
The $b$ and $c$ insertions simply absorb the  ghost zero modes.
The superghost zero modes are also absorbed by the superghost
delta-function  fields (which in the FMS formulation
correspond to ${\rm exp}(\pm \phi)$).  The only way to get a non-zero
contribution from the left and right
moving $X$ oscillators is to contract them between left and right
using
$$\langle \partial X^{\alpha} \bar \partial X^\nu \rangle =g^{\alpha \nu}{\pi
\over \tau_2}$$
where $\tau$ denotes the moduli on the torus and $\tau_2$ is its
imaginary part. The two left-over fermions are precisely
the right number needed to absorb the fermion zero modes
on the worldsheet.  Thus all the insertions absorb
zero modes of one kind or other, and we are simply left
with the partition function.  The two-dimensional matter
oscillators on the left and right cancel the (super)ghosts
on the left and right (except for the
zero mode of the bosonic matter field which gives a factor of volume),
so we are simply left with the internal
partition function for the eight dimensional theory, which is
in the (even, odd) spin structure.  This is precisely the
elliptic genus of the 8 dimensional manifold
\ref\elgen{E. Witten,``Elliptic Genera and Quantum Field Theory,''
Comm. Math. Phys. {\bf 109} (1987) 525.}\ which we denote
by $A_{M}(q)$.  Note that it is only a function of $q={\rm exp}
(2\pi i \tau)$ since the right-movers are in the odd spin structure.
Collecting all these together we thus have
$$\delta S=\epsilon^{\mu \nu}B_{\mu \nu}\cdot
{-i\over 4}\int_{\cal M}{d^2\tau \over {2 \pi\tau_2}}\cdot
{\pi \over \tau_2} A_{M}(q)$$
These are precisely the kind of objects encountered
in \lsw\ and the method for integrating over the moduli space is
also the same as used there.  We write
$${d^2\tau \over \tau_2^2}=-4i\partial \bar \partial {\rm log}(
\sqrt \tau_2 \eta \bar \eta )$$
We then use the fact
that the amplitude is total derivative in
$\bar \partial_\tau$ , which implies that we only get
the boundary contributions.   In this
limit we are left with the computation
$$\delta S={-i\over 8}
 \epsilon^{\mu \nu}B_{\mu \nu}\sum_{\partial {\cal M}}
\int d\tau_1 ({1\over  \tau_2}+{-4i\partial \eta\over \eta})A_{M}(q)$$
There are three boundaries of ${\cal M}$.  When we use modular
transformation so that they correspond to $q\rightarrow 0$,
two of them correspond to the NS sector and one
corresponds to the $R$-sector for right-movers.  It suffices
to keep the finite piece as $q \rightarrow 0$ in each of the terms.\foot{
In a similar heterotic computation we have to keep a higher order in the
$\eta$ contribution because the elliptic genus has a $q^{-1}$ term.}
In particular
$${1\over \tau_2}+{-4i\partial \eta\over \eta}\rightarrow {\pi \over 3}$$
and we replace the $A_{M}$ with the massless contribution.   In the
(NS,R) sector the massless contribution of $A_{M}$ computes the index
of the Dirac operator coupled to the tangent bundle on $M$.  Let
us call that $n_{NS,R}$. In the (R,R) sector the massless contribution
to $A_{M}$ computes minus the index of the Dirac operator coupled to the
spin bundle on $M$.
Let us call the index $n_{R,R}$.  Thus we find
$$\delta S={i\pi \over 24} \epsilon^{\mu \nu}B_{\mu \nu} (2n_{NS,R}-n_{R,R})$$
Note that the relative sign between the $n_{NS,R}$ and $n_{R,R}$ is
fixed by modular invariance, and they differ by a sign because
one is a fermionic state and the other a bosonic one.
We thus learn that
$$\int_M Y^8 \sim 2n_{NS,R}-n_{R,R}.$$
$Y^8$ can be expressed in terms of the Riemann tensor if so desired.
Actually in the context of 6d string-string duality, we are interested
in the piece in $Y^8$ which is left over after compactification on
K3.  In particular we are looking for a term in the effective
action of the form
$$\delta S = \int B \wedge {\tilde Y}^4$$
where ${\tilde Y}^4$ is the four form left after
integration ${\tilde Y}^4=\int_{K^3} Y^8$.  There is only
one combination for curvatures which involve one $\epsilon$
tensor in 4d, and that is just the Pontryagin class given by
$$p_1={1\over 16 \pi^2}\int \tr R\wedge R$$
We can fix the proportionality constant by compactifying further
to two dimensions on another $K^3$.  In this case we find that
$n_{NS,R}=-160$ and $n_{R,R}=(-16)^2=256$ which implies that
$2n_{NS,R}-n_{R,R} =-12\cdot 48=12\cdot p_1(K_3)$.

Thus we have learned\foot{ We have further
checked the absolute normalization in the above computations
using the relation between this computation and the
computation of threshold corrections
to the theta angle in Type II compactification on ${\rm K3}\times T^2$.}
that for Type IIA string
compactified
on K3 to six dimension there is a one-loop effective interaction
$$\delta S= 2\pi i \int { B}\cdot {\Theta_T\over 2}$$
where $B$ has integral periodicity
(note $B={1\over 2}\epsilon^{\mu \nu} B_{\mu \nu}$) and $\Theta_T
=\tr R\wedge R/16\pi^2$ represents the first Pontryagin class of
the six-manifold.

Note that a similar computation shows that there is no
term of the form $\int B \wedge F\wedge F$ generated at one-loop.
This follows simply because all the fundamental string states
are neutral under RR gauge fields and the corresponding index
contribution would thus vanish.  This is just as well, because
as discussed above, this term is already present at the tree
level for Type IIA strings.

The research of C. Vafa is supported in part by NSF grant PHY-92-18167;
that of E. Witten, by NSF-PHY92-45317.
\listrefs
\end